\documentclass[12pt]{article}
\usepackage{epsf,latexsym}

\newcommand{\half}{\mbox{\small $\frac{1}{2}$}}    
\newcommand{\third}{\mbox{\small $\frac{1}{3}$}}   
\newcommand{\quart}{\mbox{\small $\frac{1}{4}$}}   
\newcommand{\twothird}{\mbox{\small $\frac{2}{3}$}}
\def\lsim{\mathrel{\rlap{\lower4pt\hbox{\hskip1pt$\sim$}}
    \raise1pt\hbox{$<$}}}                
\def\gsim{\mathrel{\rlap{\lower4pt\hbox{\hskip1pt$\sim$}}
    \raise1pt\hbox{$>$}}}                

\begin{document}
\date{}

\begin{titlepage}
\renewcommand{\thefootnote}{\fnsymbol{footnote}}

\makebox[2cm]{}\\[-0.5in]
\begin{flushleft}
\begin{tabular}{l}
DESY 97-117     \\
TUM/T39-97-17   \\
HUB-EP-97/33
\end{tabular}
\end{flushleft}

\vskip0.4cm
\begin{center}

{\Large\bf A lattice determination of the second moment of the\\[6pt]
  polarised valence quark distribution\footnote{\noindent Work supported in
  part by BMBF} }\\

\vspace{2cm}

M. G\"ockeler$^a$, R. Horsley$^b$, L. Mankiewicz$^c$%
\footnote{\noindent On leave of absence from N. Copernicus
Astronomical Center, Polish Academy of Science, ul. Bartycka 18,
PL--00-716 Warsaw, Poland},
H. Perlt$^d$, P.Rakow$^e$,\\ G. Schierholz$^{e,f}$ and A. Schiller$^d$

\vspace{1.5cm}

$^a$ Institut f\"ur Theoretische Physik, Universit\"at Regensburg,  \\
     D-93040 Regensburg, Germany    \\
$^b$ Institut f\"ur Physik, Humboldt-Universit\"at zu Berlin,    \\
     Invalidenstrasse 110, D-10115 Berlin, Germany                        \\
$^c$ Institut f\"ur Theoretische Physik, TU M\"unchen, Germany            \\
$^d$ Institut f\"ur Theoretische Physik, Universit\"at Leipzig,  \\
     D-04109 Leipzig, Germany                                             \\
$^e$ Deutsches Elektron-Synchrotron DESY,                        \\
     Institut f\"ur Hochenergiephysik und HLRZ,                  \\
     D-15735 Zeuthen, Germany                                             \\
$^f$ Deutsches Elektron-Synchrotron DESY,                        \\
     D-22603 Hamburg, Germany

\vspace{1cm}

{\bf Abstract:\\[5pt]} \parbox[t] {\textwidth}{We perform a Monte
Carlo calculation of the second moment of the polarised valence quark
distributions in the nucleon, using quenched Wilson fermions.
The special feature of this moment is that it is directly accessible
experimentally. At a scale of $\mu \approx 2 \, \mbox{GeV}$ we find
$\Delta^{(1)} u = 0.198(8)$, $\Delta^{(1)} d = -0.0477(33)$. We 
compare these numbers with recent experimental results of the SMC 
collaboration.}

\end{center}
\end{titlepage}

\setcounter{footnote}{0}

\newpage

Measurements of polarised deep inelastic structure functions of the nucleon
have revealed that only a small fraction of the nucleon's spin is carried 
by the spin of the quarks~\cite{SLAC,EMC,SMC,E142,E143}. This raises the
question as to how the spin of the nucleon is distributed among spin and 
orbital angular momentum of its constituents. The constituents are
valence quarks, sea quarks and gluons. In this paper we shall be concerned 
with the spin carried by the valence quarks.

In the naive quark parton model the unpolarised structure functions and the
polarised structure function $g_1$ can be expressed in terms of the 
probability distributions for finding a quark with spin parallel, 
$q_\uparrow$, and antiparallel, $q_\downarrow$, to the longitudinally 
polarised nucleon:
\begin{eqnarray}
F_1(x) &=& \frac{1}{2} \sum_q e_q^2 [q(x) + \bar{q}(x)],\\
g_1(x) &=& \frac{1}{2} \sum_q e_q^2 [\Delta q(x) + \Delta\bar{q}(x)],
\end{eqnarray}
where
\begin{equation}
q(x) = q_\uparrow(x) + q_\downarrow(x), \;\;
\Delta q(x) = q_\uparrow(x) - q_\downarrow(x).
\end{equation}
By analogy one can define hadron inclusive structure functions
\begin{eqnarray}
F_1^h(x,z) &=& \frac{1}{2} \sum_q e_q^2 [q(x) D_q^h(z) 
+ \bar{q}(x) D_{\bar{q}}^h(z)],\\
g_1^h(x,z) &=& \frac{1}{2} \sum_q e_q^2 [\Delta q(x) D_q^h(z) 
+ \Delta\bar{q}(x) D_{\bar{q}}^h(z)],
\end{eqnarray}
where $D_q^h(z)$ is the fragmentation function for quark $q$ to produce
hadron $h$, and $z = p^h \cdot p/p \cdot q$ with $p^h$, $p$ and $q$ being 
the hadron, nucleon
and photon momentum, respectively. Of interest to us is the polarisation
asymmetry~\cite{Fra89}
\begin{equation}
A^h = \frac{g_1^h(x,z)}{F_1^h(x,z)}.
\end{equation}
Consider the polarisation asymmetry of $\pi^+$ minus $\pi^-$ inclusive 
cross sections. For a proton and deuteron target we find
\begin{equation}
A_p^{\pi^+ - \pi^-} = \frac{4 \Delta u^{\rm val}(x) - \Delta d^{\rm
val} (x)}{4 u^{\rm val}(x) - d^{\rm val}(x)}
\label{asym1}
\end{equation}
and 
\begin{equation}
A_d^{\pi^+ - \pi^-} = \frac{\Delta u^{\rm val}(x) + \Delta d^{\rm
val}(x)}{u^{\rm val}(x) + d^{\rm val}(x)},
\label{asym2}
\end{equation}
respectively. Note that the fragmentation functions drop out completely and 
that the sea quarks do not contribute. This happens because of isospin 
invariance relating the various fragmentation functions with each other.
The advantage of these quantities is that they allow a direct measurement 
of the valence quark distribution functions~\cite{Fra89}. The latter can 
then be compared with the results of 
quenched lattice calculations~\cite{goeckeler96a}.

The polarisation asymmetries (\ref{asym1}), (\ref{asym2}) have been
measured by the SMC collaboration. For the first moment
\begin{equation}
\Delta q  =  \int_0^1 dx \Delta q(x), \; q=u,d \; 
\end{equation}
it has been found, \cite{SMC96}, $\Delta u^{\rm val} = 1.01 \pm 0.19 \pm 0.14$
and $\Delta d^{\rm val} = -0.57 \pm 0.22 \pm 0.11$. The lattice results
are, \cite{dis97}, $\Delta u = 0.841(52)$ and $\Delta d = -0.245(15)$, which  
include data on $24^3 32$ lattices. 
The analysis has recently been extended to the second moment 
\begin{equation}
\Delta^{(1)} q  =  \int_0^1 dx x \Delta q(x).
\label{def1}
\end{equation}
The SMC collaboration has obtained results for $\Delta^{(1)} u$
and $\Delta^{(1)} d$~\cite{pretz97}. A
similar analysis will be published soon by the HERMES 
collaboration, \cite{Dur97}.

In QCD, $\Delta^{(1)} q = \half a_1^{(q)}$ where $a_1^{(q)}$ is the matrix 
element of the operator
\begin{equation}
   {\cal O}_5^{({\cal M})\mu \nu}
     = \left( {i\over 2} \right)
       \overline{q} \gamma^{({\cal M})\mu}\gamma^{({\cal M})5}
       \stackrel{\leftrightarrow}{D}^{({\cal M})\nu}q,
\label{mink_op}
\end{equation}
defined by
\begin{equation}
   \langle \vec{p},s|{\cal S} {\cal O}_5^{({\cal M})\mu \nu}|
   \vec{p},s \rangle 
     = a_1^{(q)} {\cal S} s^{({\cal M})\mu} p^{({\cal M})\nu},
\end{equation}
where ${\cal S}$ symmetrises and subtracts traces.
We have emphasised Minkowski space with an index ${\cal M}$.
$s^{({\cal M})}$ is the nucleon spin vector defined by
\begin{equation}
   s^{({\cal M})} = \left( {\vec{s} \cdot \vec{p} \over E}, \vec{s} \right),
   \qquad \vec{s} = \sigma m \left( \vec{e} +
                         {\vec{p} \cdot \vec{e}\over  m(E+m)} \vec{p}
                             \right),
\end{equation}
with $\vec{e}$ being the rest frame quantisation axis and
$\sigma = \pm 1$ the spin component along this axis.
We now euclideanise%
\footnote{Conventions are given in ref.~\cite{best97a}.}
and discretise the operator given in eq.~(\ref{mink_op}).
Let us first define our euclidean operator by
\begin{equation}
   {\cal O}_{5\mu\nu}
     = \half
       \overline{q} \gamma_\mu\gamma_5 
       \stackrel{\leftrightarrow}{D}_{\nu}q.
\label{eucl_op}
\end{equation}
Its transcription to the lattice is straightforward. However
on the lattice the symmetry group is reduced from $O(4)$ to the
hypercubic group $H(4)$. This loss of symmetry increases the
possibilities of mixing under renormalisation. In the continuum,
application of ${\cal S}$ is sufficient to construct operators which
are multiplicatively renormalisable (in the flavour-nonsinglet sector).
\begin{table}[ht]
    \begin{center}
        \begin{tabular}{||c|c|c|c||}
           \hline
           \multicolumn{2}{||c|}{${\cal O}(a)$}&
           \multicolumn{1}{c|}{$\langle \vec{p}, s|{\cal O}^R|
                               \vec{p}, s\rangle$}  &
           \multicolumn{1}{c||}{Representation}        \\
           \hline
           ${\cal O}_{5\{\mu\nu\}}$  & $1 \le \mu < \nu \le 4$  &
           $-a_1s_{\{\mu}p_{\nu\}}$  & $\tau_1^{(6)}$, ${\cal C}=-1$
                                                       \\
           ${\cal O}_{5,ii} -
           \quart \sum_\lambda{\cal O}_{5,\lambda\lambda}$
                                     & $1 \le i \le 3$          &
           $-a_1s_ip_i$              & $\tau_4^{(3)}$, ${\cal C}=-1$
                                                       \\
           \hline
        \end{tabular}
    \end{center}
\caption{Multiplicatively renormalisable lattice operators 
         ${\cal O}(a)$ and the continuum matrix elements of the
         corresponding renormalised operators. $\{ \cdots \}$ means
         symmetrisation. $\tau_n^{(m)}$ denotes
         the representation, \protect\cite{goeckeler96b}, with $m$ being its
         dimension. ${\cal C}$ is the charge conjugation eigenvalue of the
         operator.}
\label{table_ops}
\end{table}
To achieve the same on the lattice one has to work slightly harder
\cite{goeckeler96b}. In our case one finds two multiplets of operators
(corresponding to two inequivalent irreducible representations of
$H(4)$) which are multiplicatively renormalisable. They are given
in table~\ref{table_ops}. For each of them we obtain the renormalised
operator ${\cal O}^R (\mu)$ (renormalisation scale $\mu$) from the
lattice operator ${\cal O} (a)$ (lattice constant $a$) by multiplying
with the appropriate renormalisation factor $Z_{\cal O}$:
\begin{equation}
   {\cal O}^R(\mu) = Z_{\cal O}((a\mu)^2, g(a)) {\cal O}(a).
\end{equation}

In our computation we have only considered $\vec{p} = \vec{0}$ or
$\vec{p} = \vec{p}_1 \equiv p_1\vec{e}_1$ together with spin-quantisation
axis $\vec{e} = \vec{e}_2$. Thus from Table~\ref{table_ops} we can
consider
\begin{eqnarray}
   {\cal O}_{a_1,a} &=& {\cal O}_{5\{12\}},
                                              \nonumber \\
   {\cal O}_{a_1,b} &=& {\cal O}_{5\{42\}}.
\end{eqnarray}
Note that for a non-zero matrix element for 
${\cal O}_{a_1,a}$ we must have a non-zero three-moment\-um, while
for ${\cal O}_{a_1,b}$ we do not have this restriction.
As including three-momentum in the matrix element makes the signal
more noisy we would expect the best result to come from using zero
three-momentum together with the matrix element of ${\cal O}_{a_1,b}$. 

The nucleon matrix elements are computed from ratios of
three- to two- point functions. Thus with
\begin{eqnarray}
   C_\Gamma(t,\vec{p}) &=& \sum_{\alpha\beta} \Gamma_{\beta\alpha}
                           \langle B_\alpha(t,\vec{p})
                                   \overline{B}_\beta(0,\vec{p})
                           \rangle,
                                              \nonumber \\
   C_\Gamma(t,\tau ; \vec{p},{\cal O}_{a_1})
                       &=& \sum_{\alpha\beta} \Gamma_{\beta\alpha}
                           \langle B_\alpha(t,\vec{p}) {\cal O}_{a_1}(\tau)
                                   \overline{B}_\beta(0,\vec{p})
                           \rangle,
\end{eqnarray}
and
\begin{equation}
   R(t,\tau ; \vec{p}, {\cal O}_{a_1})
                        = {C_{\half (1+\gamma_4)i\gamma_5\gamma_2}
                             (t,\tau ; \vec{p},{\cal O}_{a_1}) \over
                          C_{\half (1+\gamma_4)}(t ; \vec{p} )},
\end{equation}
we find
\begin{eqnarray}
   R(t,\tau ; \vec{p}, {\cal O}_{a_1,a}) 
                     &=& - {1 \over Z_{{\cal O}_{a_1}}}{ 1 \over 2\kappa}
                           {1 \over 4} {m_N \over E_N} p_1 a_1
                           \qquad t \gg \tau \gg 0,   
                                              \nonumber \\
   R(t,\tau ; \vec{p}, {\cal O}_{a_1,b}) 
                     &=& - {1 \over Z_{{\cal O}_{a_1}}}{ 1 \over 2\kappa}
                           {i\over 4} m_N a_1
                           \qquad t \gg \tau \gg 0,
\label{Rratio}
\end{eqnarray}
where $\kappa$ is the Wilson hopping parameter and $m_N$ ($E_N$) is
the mass (energy) of the nucleon. Eq.~\ref{Rratio} holds provided
the operator is inserted at a Euclidean time $\tau$ sufficiently far
away from the nucleon source at time $t=0$ and the sink at time $t$.
We then expect to observe a `plateau' where $R$ is independent of $\tau$. 

\begin{figure}[ht]
\epsfxsize=15.00cm \epsfbox{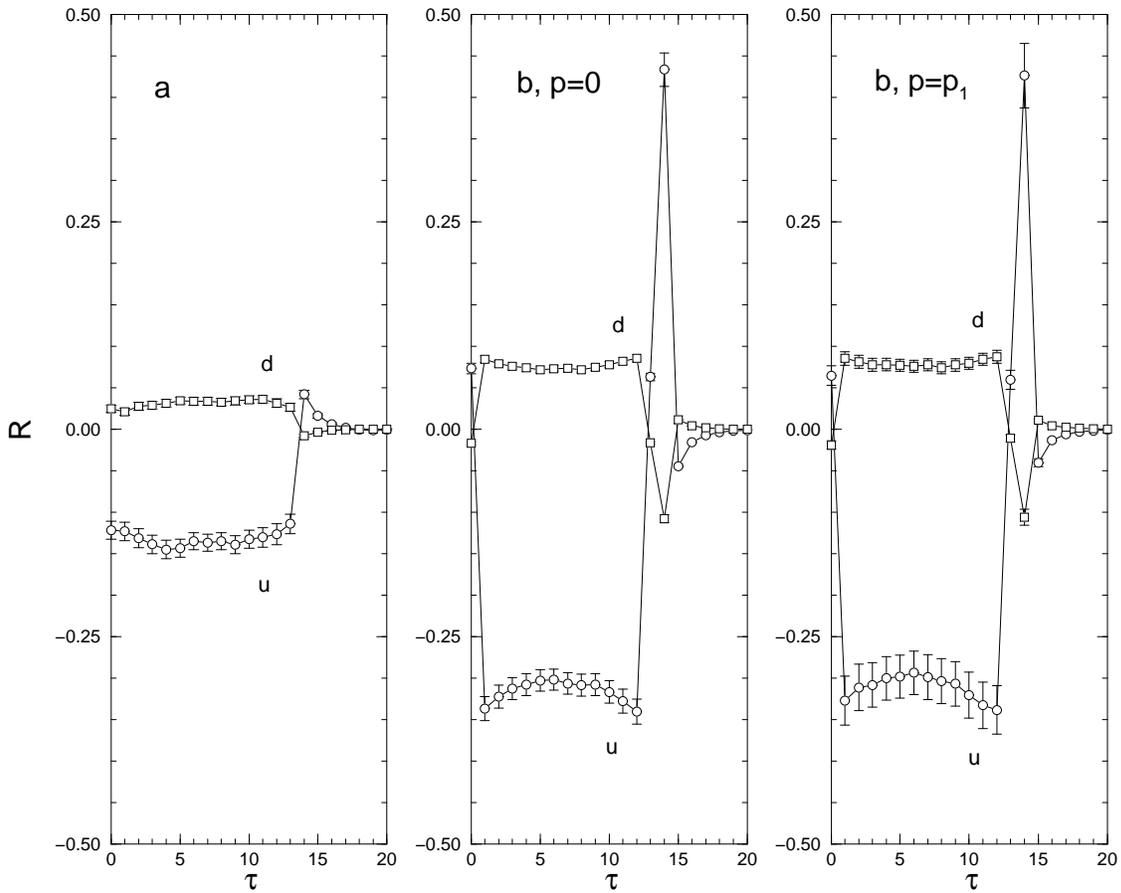}  
\caption{The $R \equiv R(t=13,\tau,\vec{p},{\cal O}_{a_1})$ ratio 
         at $\kappa=0.1530$.}
\label{fig_Rratio}
\end{figure}

We have simulated quenched Wilson fermions with five different $\kappa$
values on $N_s^3\times 32$ lattices with $N_s=16$, $N_s=24$
at $\beta \equiv 6/g^2 = 6.0$. $p_1$ is taken to be the lowest
possible momentum, namely $p_1=2\pi/N_s$.
More details of the computation may be found in ref.~\cite{goeckeler96a}.
Note, in particular, that we have dropped the quark-line disconnected
terms. This, coupled with the use of the quenched approximation,
means that sea quarks play only a small role in our calculation,
so that we are effectively measuring valence moments only.
The masses for this calculation are taken from
\cite{goeckeler97b}. In Fig.~\ref{fig_Rratio}
we show typical ratios for $t=13$. 
The fit interval for the determination of the
plateau is chosen to lie between $\tau = 4$ and $\tau=9$.
These values of $\tau$ seem to be sufficiently far away from both
the baryon source $t=0$ and sink $t=13$ to avoid contamination with
excited states.

To find $\Delta^{(1)} u$ and $\Delta^{(1)} d$, we must also calculate
the renormalisation constant $Z_{{\cal O}_{a_1}}^{\overline{MS}}$.
This was given in ref.~\cite{best97a} where it was called
$Z_{{\cal O}_{r_{2a}}}$, see also \cite{capitani97a}.
Explicitly, we have 
\begin{equation}
   Z^{\overline{MS}}_{{\cal O}_{a_1}}((a\mu)^2,g)
           = 1 - {g^2 \over 16\pi^2}C_F
                            \left[ \gamma_{{\cal O}_{a_1}} \ln (a\mu) + 
                                   B_{{\cal O}_{a_1}}^{\overline{MS}}
                            \right] + O(g^4), 
\label{z_factor}
\end{equation}
with $C_F = {4\over 3}, \gamma_{{\cal O}_{a_1}} = {16\over 3}$, and
$B_{{\cal O}_{a_1}}^{\overline{MS}}=0.3454$. We shall work, for
convenience, at a scale of $\mu = 1/a$.  This gives for $\beta = 6.0$,
$Z_{{\cal O}_{a_1}}(1,1) = 0.997$.
We see that the $O(g^2)$ coefficient is thus very small. This is because
`tadpole terms'  in the perturbation expansion have cancelled.
Indeed if we try to improve the result using
`tadpole improved boosted perturbation theory', \cite{lepage93a},
we find
\begin{equation}
   Z^{\overline{MS}}_{{\cal O}_{a_1}}((a\mu)^2,g)) =
          {u_0 \over u_0^{n_D}}
              \left( 1 - {\alpha^{\overline{MS}}(\mu) \over 4\pi} C_F
                  [\gamma_{{\cal O}_{a_1}} \ln (a\mu) +
                   B_{{\cal O}_{a_1}}^{\overline{MS}}
                                 + (n_D-1)\pi^2 ] \right),
\end{equation}
where $n_D$ is the number of derivatives,
$u_0 = \langle \third \mbox{tr} U_{\Box} \rangle^{\quart}$ 
with $U_{\Box}$ the expectation value of the plaquette, and
$\alpha^{\overline{MS}}(\mu)$ is used here as the boosted coupling.
In our case $n_D=1$, so the coefficient of the coupling constant
is unchanged and remains small, and at
$\beta = 6.0$, $\alpha^{\overline{MS}}(1/a) \approx 0.1981$,
\cite{lepage93a}, so that
\begin{equation}
   Z_{{\cal O}_{a_1}}^{\overline{MS}}(1,1) = 0.993
\label{Z_boost}
\end{equation}
which is a negligible change in the previous result.
Our conclusion is that the uncertainty in the renormalisation
constants is probably quite small, although only a full non-perturbative
calculation would reveal this of course.
We shall use, in the following, the result from eq.~(\ref{Z_boost}).

Finally we note that to determine the scale $\mu$ for our results
we must first estimate $a$. Measuring a physical quantity on the lattice
(such as a mass) and comparing this to the experimental value
determines $a$. However as Wilson fermions have $O(a)$ corrections,
it is better to set the scale from a gluonic quantity, as this has
only $O(a^2)$ corrections. This also avoids the need of making a chiral
extrapolation in the masses. From the string tension $K$ we get
(using a scale of $\sqrt{K}=427\mbox{MeV}$, \cite{eichten80a})%
\footnote{From a combined fit of lattice data, see ref.~\cite{goeckeler97b}.}
\begin{equation}
   a^{-1} = 1.95(2) \, \mbox{GeV},
\end{equation}
where the error is purely statistical.

\begin{figure}[ht]
\epsfxsize=15.00cm \epsfbox{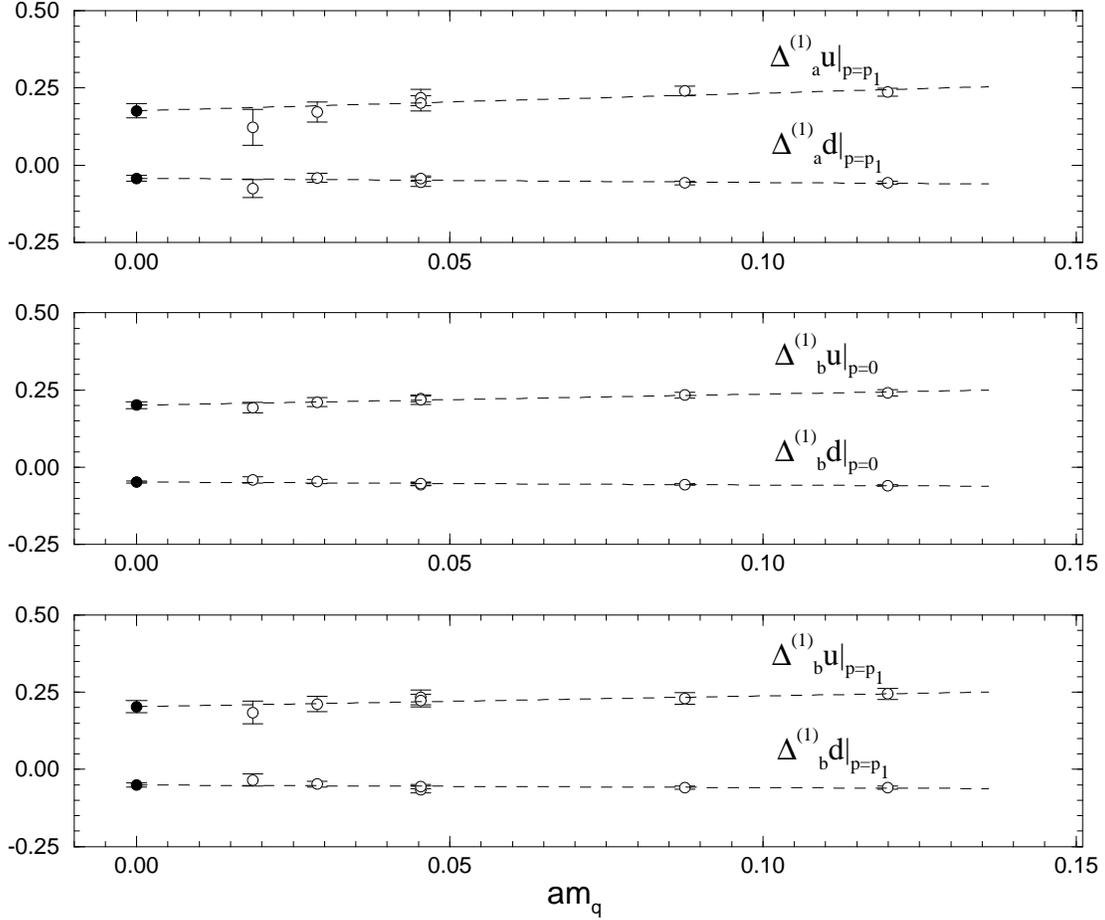}  
\caption{The chiral extrapolation for $\Delta^{(1)} q$.}
\label{fig_a1}
\end{figure}

We now come to our results. In Tables~\ref{table_res1}, \ref{table_res2}
\begin{table}[ht]
    \begin{center}
        \begin{tabular}{||l|c|c|c||}
           \hline
           \multicolumn{1}{||c|}{$\kappa$} &
           \multicolumn{1}{c|}{$0.1515$}   &
           \multicolumn{1}{c|}{$0.1530$}   &
           \multicolumn{1}{c||}{$0.1550$}  \\
           \hline
           \# configs   & $O(400)$      & $O(500)$      & $O(900)$      \\
           \hline
           $\Delta^{(1)}_a u|_{\vec{p}=\vec{p}_1}$  &
           0.236(13)    & 0.240(15)     & 0.219(26)                      \\
           $\Delta^{(1)}_b u|_{\vec{p}=\vec{0}}$    &
           0.241(10)    & 0.234(9)      & 0.223(10)                      \\
           $\Delta^{(1)}_b u|_{\vec{p}=\vec{p}_1}$  &
           0.244(18)    & 0.229(19)     & 0.233(24)                      \\
           $\Delta^{(1)}_a d|_{\vec{p}=\vec{p}_1}$  &
           -0.0577(52)  & -0.0577(60)   & -0.0549(148)                   \\
           $\Delta^{(1)}_b d|_{\vec{p}=\vec{0}}$    &
           -0.0596(27)  & -0.0559(24)   & -0.0556(38)                    \\
           $\Delta^{(1)}_b d|_{\vec{p}=\vec{p}_1}$  &
           -0.0596(48)  & -0.0586(51)   & -0.0651(10)                    \\
           \hline
        \end{tabular}
    \end{center}
\caption{The results for $\Delta^{(1)} q$ on a $16^3\times 32$ lattice.}
\label{table_res1}
\end{table}
\begin{table}[ht]
    \begin{center}
        \begin{tabular}{||l|c|c|c||}
           \hline
           \multicolumn{1}{||c|}{$\kappa$} &
           \multicolumn{1}{c|}{$0.1550$}   &
           \multicolumn{1}{c|}{$0.1558$}   &
           \multicolumn{1}{c||}{$0.1563$}  \\
           \hline
           \# configs   & $O(100)$      & $O(100)$      & $O(100)$       \\
           \hline
           $\Delta^{(1)}_a u|_{\vec{p}=\vec{p}_1}$            &
           0.200(25)    & 0.172(32)     & 0.122(58)                      \\
           $\Delta^{(1)}_b u|_{\vec{p}=\vec{0}}$              &
           0.219(14)    & 0.211(14)     & 0.193(17)                      \\
           $\Delta^{(1)}_b u|_{\vec{p}=\vec{p}_1}$            &
           0.223(21)    & 0.212(25)     & 0.184(37)                      \\
           $\Delta^{(1)}_a d|_{\vec{p}=\vec{p}_1}$            &
           -0.0437(81)  & -0.0409(141)  & -0.0755(293)                   \\
           $\Delta^{(1)}_b d|_{\vec{p}=\vec{0}}$              &
           -0.0526(48)  & -0.0463(68)   & -0.0416(103)                   \\
           $\Delta^{(1)}_b d|_{\vec{p}=\vec{p}_1}$            &
           -0.0559(63)  & -0.0473(95)   & -0.0344(195)                   \\
           \hline
        \end{tabular}
    \end{center}
\caption{The results for $\Delta^{(1)} q$ on a $24^3\times 32$ lattice.}
\label{table_res2}
\end{table}
we give our numbers and in Fig.~\ref{fig_a1}
we plot $\Delta^{(1)} u$, $\Delta^{(1)} d$ against the bare quark mass
\begin{equation}
   am_q ={1 \over 2}\left( {1\over \kappa} -  {1\over \kappa_c} \right),
\end{equation}
together with a linear chiral extrapolation. (The chiral limit
is determined from the vanishing of the pion mass, which occurs
here at $\kappa_c = 0.15721(1)$, \cite{goeckeler97b}.
We roughly estimate our quark masses from $am_q^R = Z_m am_q \sim 1.21am_q$,
\cite{goeckeler97b}, to be
$m^{\overline{MS}}(2 \, \mbox{GeV}) \sim 290$, $210$, $110$,
$70$, $45$ MeV.)
The quality of the data seems quite good.
The statistics for the light quark masses is rather low, so although
we include them in the linear fit, they should be regarded as only
confirming the heavier quark mass results.
Note that the results for $\kappa = 0.1550$ have been measured
on two different lattice sizes and differ by less than their errors
which would indicate that at least up to this value of $\kappa$ finite
size effects are under control. At each $\kappa$ we have three 
different measurements of $\Delta^{(1)}q$, one at $\vec{p}=\vec{0}$
and two at $\vec{p}=\vec{p}_1$. If Lorentz invariance has been restored,
all three measurements should agree, which is what they do within errors.
As expected the $\vec{p}=0$ results are the best,
although the other signals are quite acceptable.
It would also seem that the gradient in the chiral extrapolation
is rather small over this quark mass range. The values in the chiral
limit are given in Table~\ref{table_chiral_extrap}.
\begin{table}[ht]
    \begin{center}
        \begin{tabular}{||c|c|c|c|c|c||}
           \hline
           \multicolumn{1}{||c|}{$\Delta^{(1)}_a u|_{\vec{p}=\vec{p}_1}$}&
           \multicolumn{1}{c|}{$\Delta^{(1)}_b u|_{\vec{p}=\vec{0}}$}    &
           \multicolumn{1}{c|}{$\Delta^{(1)}_b u|_{\vec{p}=\vec{p}_1}$}  &
           \multicolumn{1}{c|}{$\Delta^{(1)}_a d|_{\vec{p}=\vec{p}_1}$}  &
           \multicolumn{1}{c|}{$\Delta^{(1)}_b d|_{\vec{p}=\vec{0}}$}    &
           \multicolumn{1}{c||}{$\Delta^{(1)}_b d|_{\vec{p}=\vec{p}_1}$} \\
           \hline
           0.176(23)  & 0.201(10)   & 0.203(20)
                      & -0.0427(94) & -0.0477(41) & -0.0507(71)    \\
           \hline
        \end{tabular}
    \end{center}
\caption{The results for $\Delta^{(1)} q$ in the chiral limit.}
\label{table_chiral_extrap}
\end{table}

For definiteness, averaging the three different measurements we
quote a result of
\begin{eqnarray}
   \Delta^{(1)} u &=& 0.198(8),             \nonumber \\
   \Delta^{(1)} d &=& -0.0477(33),
\end{eqnarray}
at a scale of $\mu^2 \approx 4$ GeV$^2$.

We first compare our result with that given in \cite{gehrmann96a}
where a phenomenological fit was made from the experimental data for
the polarised structure function $g_1$ to obtain the polarised parton
distribution functions. Taking the second moment of their $NLO$ fit
functions yields values of around $\Delta^{(1)}u \approx 0.15$,
$\Delta^{(1)}d \approx -0.055$, at a scale of $\mu^2 = 4$ GeV$^2$. 
While the $d$ moment is in agreement with our result,
their $u$ moment is a little smaller than ours.

We can see how the matrix element runs with the scale upon
using the renormalisation group equation
\begin{equation}
   \Delta^{(1)} q(\mu^\prime) =
        \left({ \alpha_{\overline{MS}}(\mu^\prime) \over 
                \alpha_{\overline{MS}}(\mu) }
                    \right)^{C_F {\gamma_{{\cal O}_{a_{1}}} \over 2b_0}}
                    \Delta^{(1)} q(\mu),
\end{equation}
with the value of $\alpha^{\overline{MS}}(2 \mbox{GeV}) \approx 0.1981$,
\cite{lepage93a}, and with $b_0=11-\twothird n_f = 11$, as we are working in
the quenched approximation. Thus at $\mu^2 = 10\mbox{GeV}^2$ 
we find:
\begin{eqnarray}
   \Delta^{(1)} u &=& 0.189 \pm 0.008,                \nonumber \\
   \Delta^{(1)} d &=& -0.0455 \pm 0.0032.
\end{eqnarray}
No great dependence on the scale is seen.

Finally our predictions can now be compared with the preliminary
results of the SMC collaboration \cite{pretz97} at the scale
$\mu^2 = 10$ GeV$^2$ which read:
\begin{eqnarray}
   \Delta^{(1)} u &=&  0.169 \pm 0.018 \pm 0.012,          \nonumber \\
   \Delta^{(1)} d &=&  -0.055 \pm 0.027 \pm 0.011,
\end{eqnarray}
where the first quoted error is statistical and the second one systematic.
The integrals have been evaluated using data from the measured range $0.003
\le x \le 0.7$. The unmeasured regions below $x = 0.003$ and above $x = 0.7$
give probably a negligible contribution to $\Delta^{(1)} u$ and $\Delta^{(1)}
d$. The lattice and experimental results agree within their respective
errors.

\section*{ACKNOWLEDGEMENTS}

The numerical calculations were performed on the APE
(Quadrics QH2) at DESY-Zeuth\-en,
with some of the earlier computations on the Bielefeld University APE.
We thank both institutions for their support. One of us (LM) was
supported by BMBF and the KBN grant 2~P03B~065~10.
MG and PR gratefully acknowledge support by the Deutsche 
Forschungsgemeinschaft.

\end{document}